\documentclass[
   ,final            
  ]
  {aipproc}

\layoutstyle{6x9}

\begin{document}

\title{New Approaches to Object Classification in Synoptic Sky Surveys}


\author{C. Donalek}{
  address={California Institute of Technology, MC 105-24, 1200 E. California Bl., Pasadena, CA 91109, USA}
}
\author{A. Mahabal}{
}
\author{S.G. Djorgovski}{
  ,
}
\author{S. Marney}{
}

\author{A. Drake}{
}
\author{E. Glikman}{
}
\author{M.J. Graham}{
}
\author{R. Williams}{
}

\begin{abstract}
Digital synoptic sky surveys pose several new object classification challenges.  In surveys where real-time detection and classification of transient events is a science driver, there is a need for an effective elimination of instrument-related artifacts which can masquerade as transient sources in the detection pipeline, e.g., unremoved large cosmic rays, saturation trails, reflections, crosstalk artifacts, etc.  We have implemented such an Artifact Filter, using a supervised neural network, for the real-time processing pipeline in the Palomar-Quest (PQ) survey. After the training phase, for each object it takes as input a set of measured morphological parameters and returns the probability of it being a real object.  Despite the relatively low number of training cases for many kinds of artifacts, the overall artifact classification rate is around 90\%, with no genuine transients misclassified during our real-time scans.  Another question is how to assign an optimal star-galaxy classification in a multi-pass survey, where seeing and other conditions change between different epochs, potentially producing inconsistent classifications for the same object.  We have implemented a star/galaxy multipass classifier that makes use of external and a priori knowledge to find the optimal classification from the individually derived ones.  Both these techniques can be applied to other, similar surveys and data sets.
 \end{abstract}

\maketitle


\section{Real-Time Classification of Astrophysical Events}

In surveys where real-time detection and classification of transient events is a science driver, there is a need for an effective elimination of instrumental data artifacts which can appear as spurious transient sources in the detection pipeline, since they are not present in the baseline comparison images.

\begin{figure}
 \includegraphics[height=.3\textheight]{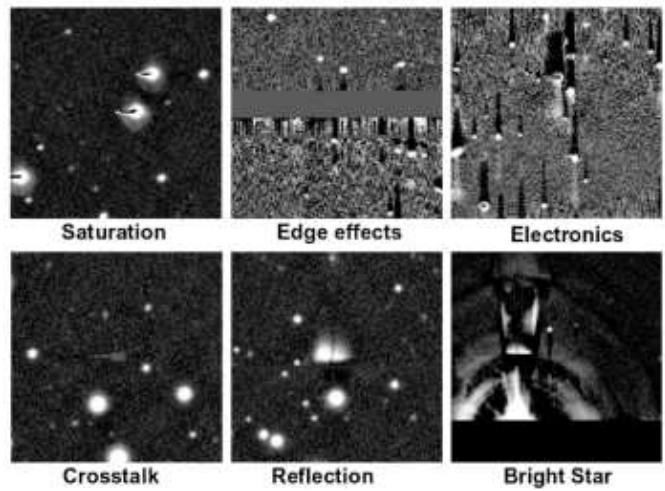}
 \caption{Image cutouts showing a variety of instrumental and data artifacts found in PQ images.}
 \label{fig:artifacts}
\end{figure}

Automated classification of candidate events, separating real astronomical sources, from a variety of instrumental and data artifacts (Fig.\ref{fig:artifacts}) has been successfully implemented within the Palomar-Quest (PQ) survey's real time data reduction pipeline and event factory \cite{Djorgovski2008} , using a set of Multi-Layer Perceptrons (MLP) trained to distinguish artifacts from genuine transients.

\begin{figure}
 \includegraphics[height=.23\textheight]{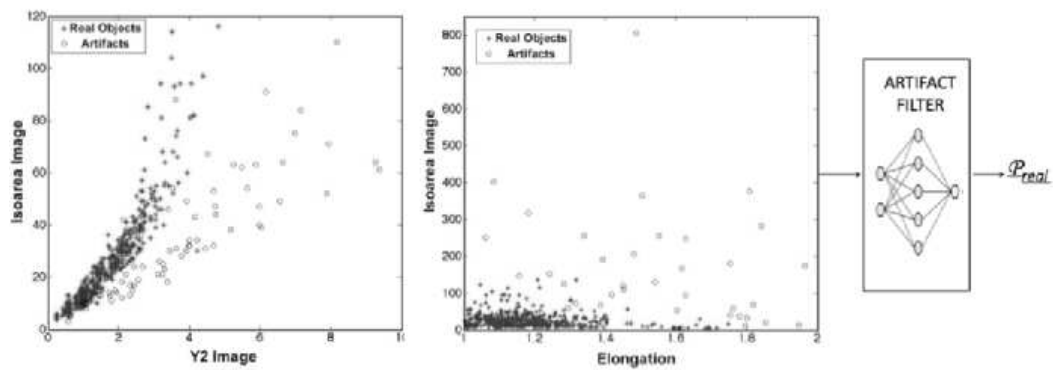}
 \caption{Artifact Filter. The two plots on the left side show a couple of morphological parameters, used to train the network, in which artifacts separate well from genuine objects.}
\label{fig:artfilter}
\end{figure}

For each potential candidate we have up to four detections, one per filter; each detection is fed separately to the artifact classifier, using as input a set of measured morphological parameters (Fig. \ref{fig:artfilter}). The outputs are then combined to have the final classification, i.e., its probability to be a real object. Adjusting thresholds allows us to be more or less aggressive discarding or keeping objects. Despite the relatively low number of training cases for many kinds of artifacts, the overall artifact classification rate is around 90\%, with no genuine transients misclassified during our real-time scans.

\section{Star/Galaxy Classification in Multi Pass Surveys}

A classical and crucial problem in the analysis of astronomical sky surveys is the morphological classification of detected sources \cite{Fayyad} \cite{Tagliaferri}. At the most basic level, this problem can be traced to the separation of sources into spatially unresolved ones (stars) and spatially resolved ones (galaxies).
With the advent of synoptic (multi-epoch) sky surveys, and the federation of multiple sky surveys and other data sets in the VO context, the problem of star-galaxy classification has come back at a much more complex level. In fact, even though a given source is necessarily  either a 'star' or a 'galaxy', when it is imaged in different conditions it can get many independent and inconsistent classifications. 

\begin{figure}[htbp]
	\centering
		\includegraphics[height=.3\textheight]{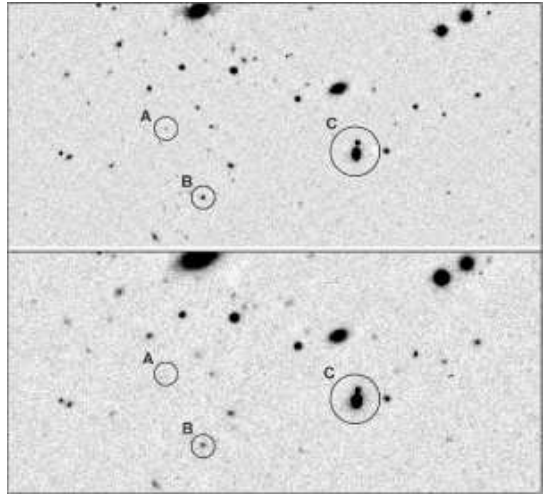}
	\caption{How the seeing quality affects the imaging passes. Upper figure: good seeing; lower figure: mediocre seeing. In the lower part many objects seem to be fuzzy, and thus potentially misclassified as galaxies (e.g. B and C) or vanishing (e.g. A).}
	\label{fig:seeingex}
\end{figure}

In addition to the information present in an image in which the sources are being first detected and then
classified, there might be some useful external, a priori information which could and should be used.
In the specific case of S/G classification, the main a priori knowledge is the seeing (Fig. \ref{fig:seeingex}). We have implemented a star/galaxy multipass classifier that makes use of external and a priori knowledge to find the optimal classification from the individually derived ones.

\subsection{Architecture of the Classifiers and Rating Criteria}

\begin{figure}
\centerline{\includegraphics[width=8cm]{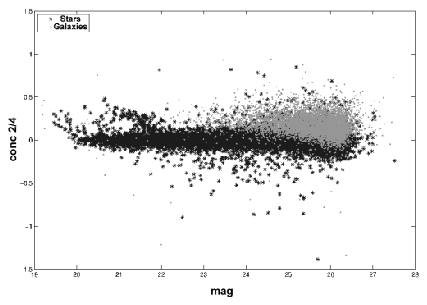}\includegraphics[width=8cm]{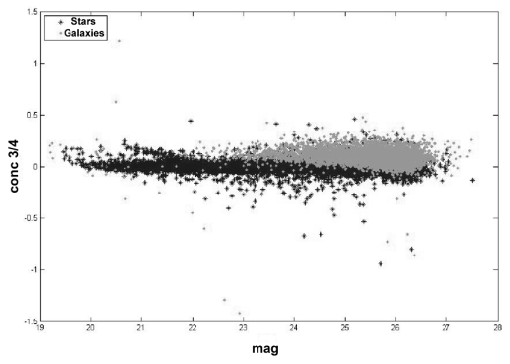}}
\caption{Two of the parameters used to train the network.}
\label{fig:parameters}
\end{figure}

All the classifiers used in the models described below, are built using a Multilayer Perceptron with a Softmax Activation
Function and Cross-Entropy Error  \cite{Bishop}, in order to interpret the network outputs as the conditional probabilities $p(C_1|\textbf{x})$ and $p(C_2|\textbf{x})$ where \textbf{x} is the input vector, $C_1$ the galaxy class, $C_2$ the star class. The input pattern fed to the MLP consist of a set of image 
parameters (Fig. \ref{fig:parameters}), extracted from the images processed 
by the PQ pipeline and overlapped with the 
SDSS database \cite{SDSSwebsite} that gives us the true class of each object.
The network output can be viewed as the "stellarity", i.e., how much a given object can be considered a star. This kind of analysis is useful when some mistakes can be more costly than others. 

The performances of the classifiers are rated based on the following three criteria: completeness (eg., the percentage of stars correctly classified as such), contamination (eg., the percentage of stars  incorrectly classified as galaxies); overall classification rate (percentage of objects correctly classified).

\subsection{Different Approaches of Introducing A Priori Knowledge}

\begin{figure}
\centering
\includegraphics[width=4cm]{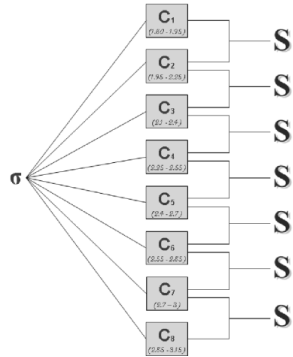}
\caption{Classification tree with overlap.}
\label{fig:overlap}
\end{figure}

In principle, introducing a priori knowledge leads to faster learning 
speed and better generalization ability. Introducing the seeing directly in the input to the network
does not improve the performance.
However, it is often found that improved performance can be obtained by combining
models together in some way, instead of using a single model in isolation.
In this way, individual classifiers may be optimized or trained differently. In the classification tree the individual models are generally choosen to be very simple and the overall flexibility of the model arises from the input-dependant selection process. Our classification tree has been built 
using the seeing ($\sigma$) as decision flag and training 4 different 
MLPs, each one responsible for a given seeing range. One main limitation of the classification trees is that they partition the input space in regions with hard boundaries in which only one model is responsible for making predictions for any given value in input. In order to overcome such problem we introduced an overlapping structure (Fig. \ref{fig:overlap}) \cite{Donalek}, in such a way that each datum falls in at least two classifier ranges and the final classification can be derived, for example, from the weighted mean of the individual results.

Table \ref{table:startoc} shows the result of the classification tree with overlap. 
As it can be easily seen, performances at higher values of seeing are by far improved with 
respect to those of the classifier without a priori knowledge. Using a classification
tree with overlap and a stellarity threshold greater than 0.90 (Table \ref{table:starcc}), the star contamination
is always very low, even at higher seeing where we have $6.4\%$ for the simple classifier 
against only $\sim 2\%$ of the classification tree with overlap.
At lower values of seeing ($\sigma < 2.7$), 
the contamination is low for all the models, but we note an higher completeness for the classification trees, expecially for the one with overlap.
\\

\begin{table}[h]
\centering
\label{table:startoc}
\smallskip
\begin{tabular}{cccc}  \hline  \hline
object & C & CTO & $\sigma$ \\ \hline
Star (objid=1185) & 99.25\% & 99.67\% & $2.21$ \\
Star (objid=1185) & 97.98\% & 98.80\% & $2.28$ \\
Star (objid=1185) & 95.44\% & 95.78\% & $2.67$ \\
Star (objid=1185) & 82.79\% & 90.38\% & $3.67$ \\ 
Star (objid=1722) & 89.00\% & 90.50\% & $2.34$ \\ 
Star (objid=1722) & 96.14\% & 94.38\% & $2.50$ \\ 
Star (objid=1722) & 78.20\% & 90.10\% & $3.27$ \\ \hline
\end{tabular}
\caption{How the same stars, detected in multiple passes, are classified using a classifier without a priori knowledge (C) and the classification tree with overlap (CTO).}\end{table}
\smallskip

\begin{table}[h]
\centering
\caption{Star completeness and contamination with stellarity threshold $= 0.90$.}
\label{table:starcc}
\smallskip
\begin{tabular}{ccccccc}  \hline  \hline
  & C1 & CT & CT ov. & C1 & CT & CT ov. \\ \hline
seeing & Compl. & Compl. & Compl. &  Cont. & Cont. & Cont. \\
$1.80<\sigma<1.95$ & 82.16\% & 89.58\% & 89.58\% & 1.95\% & 1.79\% & 1.79\% \\
$1.95<\sigma<2.10$ & 82.68\% & 87.83\% & 86.46\% & 1.45\% & 1.50\% & 1.50\% \\
$2.10<\sigma<2.25$ & 86.83\% & 88.49\% & 90.54\% & 1.10\% & 1.14\% & 1.12\% \\
$2.25<\sigma<2.40$ & 80.10\% & 82.10\% & 83.83\% & 1.46\% & 1.55\% & 1.33\% \\
$2.40<\sigma<2.55$ & 80.86\% & 82.62\% & 84.80\% & 1.53\% & 1.42\% & 1.40\% \\
$2.55<\sigma<2.70$ & 81.50\% & 82.00\% & 82.50\% & 2.36\% & 2.27\% & 2.10\% \\
$2.70<\sigma<2.85$ & 81.60\% & 81.00\% & 81.50\% & 2.47\% & 2.11\% & 1.91\% \\
$2.85<\sigma<3.00$ & 82.50\% & 81.14\% & 81.44\% & 3.50\% & 3.20\% & 1.70\% \\
$3.00<\sigma<3.15$ & 71.50\% & 74.33\% & 73.36\% & 6.4\% & 2.38\% & 2.02\% \\ \hline
\end{tabular}
\end{table}


\begin{theacknowledgments}
This work was supported in part by the NSF grants AST-0407448, AST-0326524, CNS-0540369, and by the Ajax Foundation.  S.M. was supported in part by a Caltech SURF program.  We are thankful to numerous collaborators, and especially the PQ survey team.
\end{theacknowledgments}

\end{document}